\documentclass[reprint,amsmath,amssymb,aps,prl,showpacs,superscriptaddress,showkeys,floatfix]{revtex4-1}

\usepackage{graphicx}
\usepackage{rotating}
\usepackage{amsmath}
\usepackage{bm}
\usepackage{subfigure}
\usepackage{color}
\usepackage{natbib}
\usepackage{longtable}
\usepackage{dcolumn}
\usepackage[usenames,dvipsnames]{xcolor}

\begin{document}
\widetext

\title{Density-functional Theory for $f$ electron Systems: the $\alpha$-$\gamma$ Phase Transition in Cerium}

\author{Marco Casadei}
\affiliation{Fritz-Haber-Institut der Max-Planck-Gesellschaft, Faradayweg 4-6, D-14195 Berlin, Germany} 
\author{Xinguo Ren}
\affiliation{Fritz-Haber-Institut der Max-Planck-Gesellschaft, Faradayweg 4-6, D-14195 Berlin, Germany} 
\author{Patrick Rinke}
\affiliation{Fritz-Haber-Institut der Max-Planck-Gesellschaft, Faradayweg 4-6, D-14195 Berlin, Germany} 
\author{Angel Rubio} 
\affiliation{Fritz-Haber-Institut der Max-Planck-Gesellschaft, Faradayweg 4-6, D-14195 Berlin, Germany} 
\affiliation{Nano-Bio Spectroscopy group, Dpto. Fısica de Materiales, Universidad del Pa\'is Vasco, Centro de Fısica de Materiales CSIC-UPV/EHU-MPC and DIPC, Av. Tolosa 72, E-20018 San Sebastian, Spain} 
\author{Matthias Scheffler}  
\affiliation{Fritz-Haber-Institut der Max-Planck-Gesellschaft, Faradayweg 4-6, D-14195 Berlin, Germany}

\date{\today}

\begin{abstract}
The isostructural $\alpha$-$\gamma$ phase transition in cerium is analyzed using density-functional theory with different exchange-correlation functionals, in particular the PBE0 hybrid functional and the exact-exchange plus correlation in the random-phase approximation [(EX+cRPA)@PBE0] approach. We show that the Hartree-Fock exchange part of the hybrid functional actuates two distinct solutions at zero temperature that can be associated with the $\alpha$ and $\gamma$ phases of cerium. However, despite the relatively good structural and magnetic properties, PBE0 predicts the $\gamma$ phase to be the stable phase at ambient pressure and zero temperature, in contradiction with low temperature experiments. EX+cRPA reverses the energetic ordering, which emphasizes the importance of correlation for rare-earth systems.
\end{abstract}

\maketitle

The first-principles description of $f$ electron materials is a considerable challenge and a highly debated topic in condensed-matter physics. The simultaneous presence of itinerant \emph{spd} states and localized partially occupied $f$ states and their mutual interaction in rare-earth materials give rise to a rich variety of physical phenomena that continue to be a testing ground for electronic-structure theories. Cerium is one of the most prominent representatives in this regard and, even more intriguingly, undergoes an isostructural (fcc) $\alpha$-$\gamma$ phase transition  accompanied by a  volume collapse of 15\% at room temperature and ambient pressure \cite{D.C.KoskenmakiandK.A.Gschneidner1978,Lipp2008}. The $\alpha$ phase is characterized by enhanced Pauli paramagnetism and has a smaller volume, while the larger-volume $\gamma$ phase follows a Curie-Weiss behavior for the magnetic susceptibility. 

At zero temperature, first-principles calculations have so far been unable to produce a double minimum in the total energy versus volume curve, that would be a direct indication of the phase transition, within a single theoretical framework.  In local or semilocal (LDA or GGA) functionals of density-functional theory (DFT) the $f$ electrons are always delocalized, due to the strong self-interaction error of the functionals, and only the $\alpha$ phase is described with some confidence \cite{Luders2005,Wang2008}.  The self-interaction corrected local spin density approximation \cite{Szotek1994,Svane1994,Luders2005} and Hubbard $U$ augmented local or semilocal DFT calculations  (LDA/GGA+U) \cite{Shick2001,Wang2008} enforce localization of the $f$ electrons. They subsequently yield a phase, whose volume and magnetic moment are consistent with the $\gamma$ phase, but the description of the $\alpha$ phase requires a different treatment, namely LDA. Dynamical mean field theory in combination with LDA (LDA+DMFT) has been applied to the study of the phase transition at finite temperatures \cite{Held2001,McMahan2003,Amadon2006,Haule2005}, but could so far not be extended to the zero temperature limit. Whether a double minimum exists or would only emerge in the free energy curve at finite temperature, due to entropic effects as suggested by Amadon \emph{et al.} \cite{Amadon2006}, is therefore still a matter of debate. 

In this Letter we show that hybrid density functionals \cite{Becke1993,PBE0_1,HSE_1}, that incorporate a fraction of exact exchange 
yield  a double minimum within a single theoretical and computational framework. The results are further improved quantitatively by employing exact exchange plus correlation in the \emph{random-phase approximation} (EX+cRPA) (see Ren \emph{et al.}  \cite{Ren2012b} and references therein)\footnote{We regard EX+cRPA as our \emph{benchmark}. The recent beyond RPA schemes second order screened exchange and single excitation corrections appear very promising, but their  performance for metallic systems has not been established yet and we therefore refrain from using them in this work.}. In our approach all electrons are treated on the same quantum mechanical level, in contrast to LDA/GGA+$U$ or LDA+DMFT studies.
We obtain two distinctly different solutions, whose structural, electronic and magnetic properties are consistent with experimental results for the $\alpha$ and $\gamma$ phase, respectively. 

\begin{figure}[t]
\includegraphics[width=0.45\textwidth]{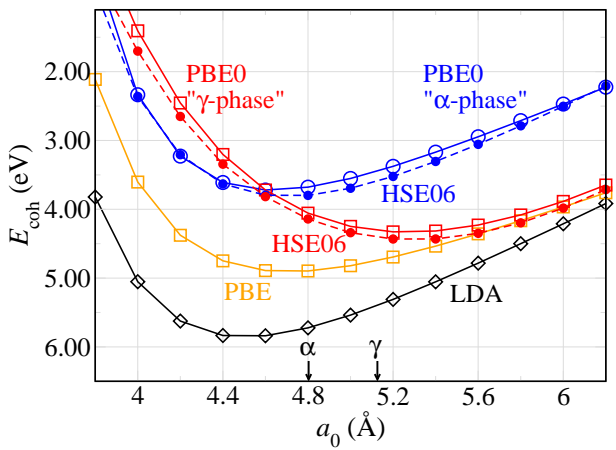}
\caption{\label{fig:pbe0endos} Cohesive energy [$E_{\mathrm{coh}}=-(E-\sum E^{\mathrm{atom}})$] of cerium as a function of the lattice constant ($a_0$). Dashed lines show HSE06 results. The spin moment increases with volume for the LDA and PBE solutions, while in PBE0 and HSE06 it remains approximately constant to zero and one-half for the $\alpha$ and $\gamma$ phases, respectively. Experimental lattice parameters for the two  phases at finite temperature \cite{D.C.KoskenmakiandK.A.Gschneidner1978}  are marked by black arrows.}
\end{figure}

All calculations in this work were performed with the all-electron code {\footnotesize FHI-AIMS} (Fritz-Haber-Institut \emph{ab initio molecular simulation}) \cite{Blum2009,Ren2012a}, that is based on numeric atom-centered orbitals. Relativistic effects are treated at the level of the scaled zero-order regular approximation \cite{Lenthe/Baerends/Snijders:1994}. Here we present results obtained using the PBE0 hybrid functional \cite{PBE0_1} for both cluster and periodic systems \cite{Levchenko2012} and show that the HSE hybrid functional \cite{HSE_1} yields a similar description. For comparison we also applied the local-density approximation in the parameterization of Perdew and Zunger \cite{Perdew/Zunger:1981} and the Perdew-Burke-Ernzerhof  generalized gradient approximation (PBE) \cite{Perdew1996}. Periodic calculations were performed with a $6\times6\times6$ $k$ mesh. 
This gives energies that are converged to within 5 meV, which is sufficient for the energy scale of interest here [cf. Fig.~\ref{fig:pbe0endos} (c)]. The hybrid functional calculations were carried out with a tier 1 numeric atom-centered orbitals  basis \cite{Blum2009}, whereas for (EX+cRPA)@PBE0 it proved necessary to go up to tier 3 \footnote{{\footnotesize FHI-AIMS} NAO basis sets for Cerium: \emph{tier}~1: [Xe]+$2s2p2d2f1g$; \emph{tier}~2: [Xe]+$3s3p4d3f2g1h$; and
\emph{tier}~3: [Xe]+$4s4p5d5f3g2h$. }. In both cases ferromagnetic ordering is assumed in our spin-polarized calculations. 

In cerium the 4$f$, 5$d$, and 6$s$ states all lie in the vicinity of the Fermi energy, giving rise to different electronic configurations that are very close in energy. In cases like this, approaches that are based on the density matrix rather than the density, such as DFT+$U$, hybrid functionals or Hartree-Fock are more susceptible to local minima in the potential-energy landscape of the electrons \cite{Fukutome1971,Mundim1996,Shick01,deAndrade/etal:2005,Thom2008,Jollet2009,Meredig10}. 
This fact can be exploited to search for distinct, stable electronic configurations. 
For a given lattice constant $a_0$ our PBE0 calculations are initialized with the density matrix from a preceding PBE calculation and with a high electronic temperature ($T\sim$1000 K). The electronic temperature yields a broadening of the one-electron energy levels. 
In subsequent calculations the temperature is gradually reduced until a particular solution is stabilized at $T=0$. When scanning a range of lattice constants ($a_0$) for cerium metal, the binding energies from different values of $a_0$ fall into two different smooth curves. 
In practical calculations, the solution at one particular lattice constant can be used to initialize the calculations at a neighboring lattice constant. In this way, the two binding energy curves can be stabilized with relative ease.

It is one of the core results of this work that the above technique provides two different stable solutions.  In Fig. \ref{fig:pbe0endos} the cohesive energies ($E_{\mathrm{coh}}$) obtained from LDA, PBE, PBE0, and HSE06 are presented as a function of the lattice constant. Our LDA and PBE results are in agreement with previous calculations \cite{Luders2005,Wang2008} and exhibit only one minimum. The associated volume is consistent with the $\alpha$ phase, although the actual value is underestimated. Constraining the magnetic moment does not introduce a second minimum. In contrast, in PBE0 and HSE06 two stable solutions are found. One solution has a minimum approximately coinciding with the LDA or PBE minimum, while the second assumes its minimum at a much larger lattice constant, consistent with the one of the $\gamma$ phase. The magnitude of the cohesive energy systematically reduces from LDA to PBE, and to PBE0.

\begin{figure}[b]
\includegraphics[width=0.45\textwidth]{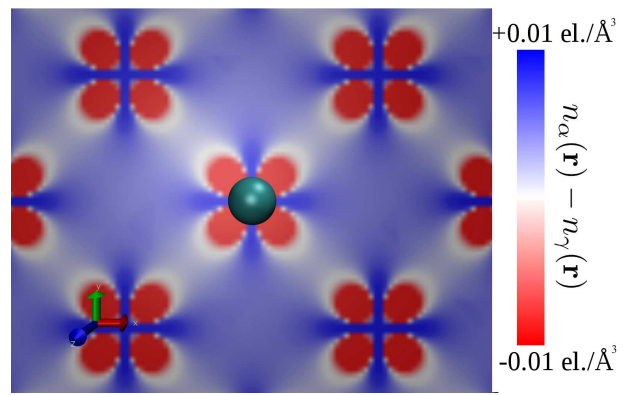}
\caption{\label{fig:densitydiff} Volume slice through the difference between the bulk Ce electron densities of the $\alpha$ and $\gamma$ phases at the same lattice constant of 4.6 \AA, at which both phases have the same energy (see Fig. \ref{fig:pbe0endos}). The $\alpha$ phase has a larger contribution in the interstitial region, whereas the $\gamma$ phase is more localized and exhibits  a clear $f$-orbital shape.}
\end{figure}

\begin{figure}[t]
\includegraphics[width=0.4\textwidth]{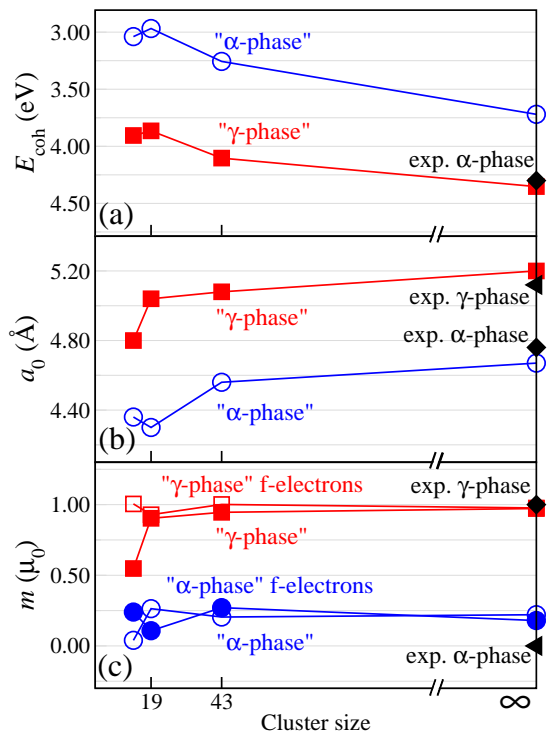}
\caption{\label{fig:extrapolation} PBE0 results for cerium clusters and corresponding bulk values for (a) the cohesive energy ($E_{\mathrm{coh}}$), (b) the lattice constant ($a_0$) and (c) the magnetic moment on the central atom ($m$). All clusters exhibit two solutions that converge to the calculated bulk limit. Experimental results marked on the right axis are taken from Ref. \cite{D.C.KoskenmakiandK.A.Gschneidner1978}.} 
\end{figure}

The two PBE0 solutions differ in their electronic structure as e.g. reflected in the density of states (see Ref. \cite{casadei2012}) 
and the magnetic moment $m$ (cf Fig. \ref{fig:extrapolation} (c)). $m$ of the low volume phase lies around 0.2 $\mu_0$, while in the high volume phase $m$ is close to one. A rapid change of the local magnetic moment across the $\alpha$-$\gamma$ phase transition was also observed in LDA+DMFT \cite{Held2001}.  Also the number of $f$ electrons is approximately one in both phases, as suggested by positron annihilation experiments \cite{Gustafson1969}. Figure \ref{fig:densitydiff} shows the difference of the PBE0 electron density between the $\alpha$- and the $\gamma$-like solutions, $n_{\alpha}(${\bfseries r}$)-n_{\gamma}(${\bfseries r}$)$, projected onto a volume slice at the same lattice constant of 4.6 \AA. The $\alpha$-like phase has a higher density in the interstitial region, a clear indication of electron delocalization, whereas in the $\gamma$-phase solution the electron is more localized around the Ce atom. More importantly, however, the density difference has the shape of an $f$ orbital of $xyz, z(x^2-y^2)$ symmetry, as evidenced by its three-dimensional projection (not shown). This provides a strong indication that the balance between localization and delocalization of the $f$ electrons plays a key role in the emergence of the double minimum in the cohesive energy curve. 

Further inspection of Fig. \ref{fig:pbe0endos} reveals that the energetic ordering of the two PBE0 solutions is not consistent with the experimental phase diagram \cite{Decremps2011}, which  shows that the $\alpha$ phase is lower in energy than the $\gamma$ phase at low temperatures. The opposite is true for the PBE0 results. To overcome this discrepancy we resort to a more accurate treatment of exchange and correlation. As the name indicates, within EX+cRPA the exchange term is treated exactly (i.e. not reduced by a factor as in hybrid functionals), and correlation is treated at the level of the \textit{random-phase approximation}. The mixing factor in the hybrid functionals that controls the fraction of exact exchange may be regarded as a simplified screening function, which is replaced by a physical and system-dependent screening in EX+cRPA. EX+cRPA is so far only implemented for finite systems in the current version of {\footnotesize FHI-AIMS}. We therefore adopt the strategy of modeling bulk Ce with systematically increasing cerium clusters. The clusters are cut from the fcc crystal structure, with one atom in the center surrounded by shells of first, second, and third nearest neighbors (i.e. thirteen, nineteen, and forty-three atoms in total). To reduce edge effects, we use the formula for evaluating the effective cohesive energy of clusters (see, e.g., \cite{SchefflerM.andStampflC.,Hu2007}) 
\begin{equation}
E_{\mathrm{coh}}=-\left[E-\sum_{c=1}^{12}(N_cE_c^{\mathrm{atom}})\right] 
         \left(\sum_{c=1}^{12}N_c\sqrt{\frac{c}{12}} \right)^{-1} 
\end{equation}
where $E$ is the total energy, $N_c$ the number of atoms in the cluster with $c$ nearest neighbors, and $E_c^{\mathrm{atom}}$ is the atomic total energy for a $c$-fold-coordinated atom. Basis-set superposition errors are corrected when evaluating $E_c^{\mathrm{atom}}$.  For each cluster two distinct solutions were always found. Figure \ref{fig:extrapolation} demonstrates how the two sets of  PBE0 cohesive energies, equilibrium lattice constants, and magnetic moments (on the central atom) of the clusters converge towards the corresponding bulk values as the cluster size grows. It can be seen from Fig. \ref{fig:extrapolation} that the essential physics is already captured with the smallest cluster size.  
In the following, the (EX+cRPA)@PBE0 results for the Ce$_{19}$ cluster will be shown, which suffices for the discussion below.

\begin{figure}[b]
\vspace{0.1cm}
\includegraphics[width=0.45\textwidth]{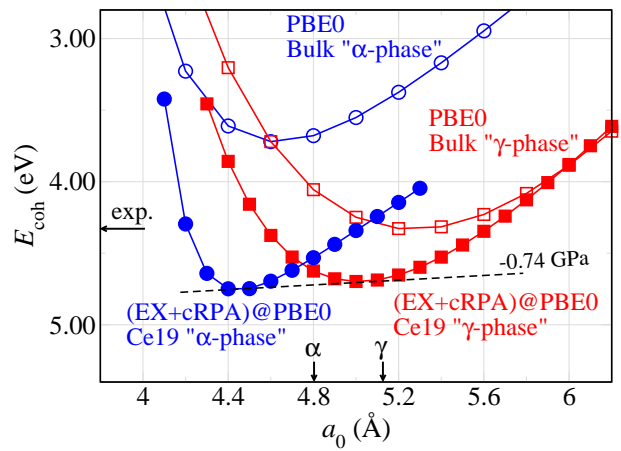}
\caption{\label{fig:rpatoten} Calculated (EX+cRPA)@PBE0 cohesive energy ($E_{\mathrm{coh}}$) for the 19-atom fcc-cerium cluster as a function of $a_0$. The dashed line illustrates the Gibbs construction for the transition pressure in good agreement with the extrapolated experimental  $P_t \simeq -0.8$ GPa (\cite{Decremps2011}). Arrows on the energy axes: experimental cohesive energy from \cite{Kittel1997}.}
\end{figure}

The EX+cRPA calculations are performed non-self-consistently with the orbitals and eigenenergies of the two PBE0 solutions as input.
Figure \ref{fig:rpatoten} shows the (EX+cRPA)@PBE0 cohesive energy for the 19-atom cluster.
Since the density of states near the Fermi level is higher in the low-volume phase, and consequently screening and the RPA correlation energy are larger, the low-volume phase moves down in energy relative to the high-volume one and the correct energetic order is restored. According to the extrapolation of the experimental data  \cite{Amadon2006,Decremps2011}, the difference in internal energy ($\Delta U$) between the two phases should lie between 20 and 30 meV. The (EX+cRPA)@PBE0 value for the 19-atom cluster amounts to $\Delta U\simeq45$ meV. Although the difference is larger than the experimental estimation of the maximum energy difference between the two phases, it is comparable to the entropy contribution $T\Delta S$ at ambient conditions \cite{Decremps2011}. Therefore, our results on the electronic contribution to the phase transition do not rule out  that entropy might play a noticeable role in the phase transition at finite temperature as proposed in Ref. \onlinecite{Amadon2006}. 
The common tangent construction to the (EX+cRPA)@PBE0 cohesive energy curves leads to a transition pressure of $P_t\simeq-0.74$ GPa at zero temperature, in good agreement with the extrapolated experimental $P_t\simeq-0.8$ GPa. The calculated lattice constants for the $\alpha$- and $\gamma$-like phases are $4.45$ and $5.03$ \AA, respectively. This corresponds to  a volume collapse of $\simeq30\%$ at zero temperature, to be contrasted with the 15\% observed experimentally at ambient conditions \cite{D.C.KoskenmakiandK.A.Gschneidner1978}.

Over the years many experimental and theoretical studies have addressed the origin of the transition but a conclusive answer is still lacking. The two prevalent propositions are: a Mott transition for the $f$ electrons \cite{Johansson1974,Johansson1995} and the Kondo volume collapse \cite{Allen1982,Allen1992}. In the Mott picture the hybridization between $f$ orbitals is believed to change  across the transition, leading to one phase with delocalized $f$ electrons ($\alpha$ phase) and the other with localized $f$ states ($\gamma$ phase). In the Kondo volume collapse model, the $f$ electrons are assumed to be localized in both phases, and the change in spin screening of the localized moments by the conduction \emph{spd} electrons is responsible for a change in the Kondo temperature across two orders of magnitude, with an associated change in system properties.  Although the two models are different in nature, recent theoretical works suggest that the resulting scenarios are quite similar at finite temperature \cite{Held2000} and both are consistent with available experimental data \cite{Lipp2008,Johansson2009}. The common belief that the phase transition should be driven by changes in the electronic structure alone has also been questioned, but the contribution of e.g. lattice vibrations has been estimated to be lower than 30\% of the total energy change  \cite{Manley2003,Decremps2009,Krisch2011}. 
Since our calculations are performed at zero temperature they can only describe a pressure induced phase transition at zero temperature but not a temperature induced transition at ambient pressure as observed in the experimental phase diagram. However, the occurrence of a double minimum in the cohesive energy curve and the strong signature of $f$ states in the accompanying density difference shown in  Fig. \ref{fig:densitydiff} suggest that the localization of $f$ electrons is a strong contributing factor to the phase transition. Density-functional theory (being a ground state theory) does not need to give the right electronic excitation spectrum to describe the structural phase transition at zero temperature, even if the spectrum is essential for the distinction between the Mott transition and the Kondo volume collapse at finite temperature. Nevertheless, our zero temperature results are consistent with the scenario of a Mott transition and show that advanced DFT exchange-correlation functionals can indeed capture such Mott physics.

In conclusion, PBE0 hybrid functional calculations combined  with EX+cRPA produce multiple distinct solutions of the electronic structure of bulk cerium at zero temperature. These can be discriminated by their magnetic moment and the degree of $f$ electron localization and have been tentatively associated with the $\alpha$ and $\gamma$ phases of cerium. At the PBE0 level, the energetic ordering of the two solutions is reversed compared to the zero temperature extrapolation of the experimental phase diagram. EX+cRPA recovers the right ordering, which highlights the role of correlation in rare-earth systems. 
 
An interesting aspect that emerged from the cluster extrapolation approach is the presence of a volume collapse at the nanoscale down to the dimer. This is a very unusual feature for a first-order phase transition in a metal and opens the question of whether other lanthanides would exhibit the same behavior. This prediction suggests further investigation both theoretically and experimentally. 

\begin{acknowledgements}
A. R. acknowledges financial support from the European Research Council (ERC-2010-AdG -No. 267374), Spanish (FIS2011-65702-C02-01) and Grupos Consolidados UPV/EHU del Gobierno Vasco (IT-319-07).
\end{acknowledgements}


%

\end{document}